%% file: main.tex
\documentclass[journal]{IEEEtran}
\usepackage{amsmath}
\usepackage{bbm}
\usepackage{graphicx}
\usepackage{xcolor}
\PassOptionsToPackage{hyphens}{url}\usepackage{hyperref}
\usepackage{amsfonts}
\usepackage{algorithm,algorithmic}
\begin{document}

\makeatletter
\def\footnoterule{\kern-3\p@
  \hrule \@width 2in \kern 2.6\p@} 
\makeatother

\title{Enhancing Video Streaming in Vehicular Networks via Resource Slicing}

\author{Hamza Khan,
        Sumudu~Samarakoon, \IEEEmembership{Member, IEEE},
        and Mehdi Bennis, \IEEEmembership{Senior~Member, IEEE}%
        
    \thanks{Hamza Khan (Corresponding author), and Sumudu Samarakoon are with the Centre for Wireless Communications, University of Oulu, 90014 Oulu, Finland (emails: \{hamza.khan,\,sumudu.samarakoon\}@oulu.fi).}
     \thanks{ Mehdi~Bennis is with the Centre for Wireless Communications, University of Oulu, 90014 Oulu, Finland, and also with the Department of Computer Science and Engineering, Kyung Hee University, Seoul 17104, South Korea (e-mail: mehdi.bennis@oulu.fi).}

\vspace{-0.4cm}}

\maketitle

\begin{abstract}
\input{abstract.tex}
\end{abstract}

\IEEEpeerreviewmaketitle
\section{Introduction}
\input{intro.tex}

\section{System Model and Problem Definition}
\input{system_model.tex}

\section{Vehicle Clustering Via Network Slicing}
\input{vehicle_clustering.tex}

\section{Vehicle scheduling and video selection via Lyapunov framework}
\input{lyapunov.tex}

\section{Performance Analysis}
\input{performance_analysis.tex}

\section{Conclusion}
\input{conclusion.tex}

\section*{Acknowledgment}
\input{ack.tex}

\bibliographystyle{IEEEtran}
\bibliography{main}

\end{document}

%% file: abstract.tex
Vehicle-to-everything (V2X) communication is a key enabler that connects vehicles to  neighboring vehicles, infrastructure and pedestrians. In the past few years, multimedia services have seen an enormous growth and it is expected to increase as more devices will utilize infotainment services in the future i.e. vehicular devices. 
Therefore, it is important to focus on user centric measures i.e. quality-of-experience (QoE) such as video quality (resolution) and fluctuations therein.  
In this paper, a novel joint video quality selection and resource allocation technique is proposed for increasing the QoE of vehicular devices. The proposed approach exploits the queuing dynamics and channel states of vehicular devices, to maximize the QoE while ensuring seamless video playback at the end users with high probability. The network wide QoE maximization problem is decoupled into two subparts. First, a network slicing based clustering algorithm is applied to partition the vehicles into multiple logical networks. Secondly, vehicle scheduling and quality selection is formulated as a stochastic optimization problem which is solved using the Lyapunov drift plus penalty method. Numerical results show that the proposed algorithm ensures high video quality experience compared to the baseline. Simulation results also show that the proposed technique achieves low latency and high-reliability communication.

%% file: intro.tex
In the recent years, multimedia services have seen exponential growth and their demand is increasing every day. In 2015 alone, multimedia traffic was more than half of the global mobile data traffic \cite{cisco}. Tremendous growth in traffic is one of the many driving forces behind the next generation of mobile services (5G) which is expected to serve a vast variety of connected devices with several use cases, i.e. enhanced mobile broadband (eMBB), ultra-reliable low latency communication (URLLC) and massive machine type communication (mMTC). 
Vehicles are the fastest growing type of connected devices after cellphones and tablets \cite{petri}. Vehicle-to-everything (V2X) communication has been studied from over a decade due to its potential in enabling safer transportation. Apart from safety application, vehicular infotainment services are getting more attention mainly due to multimedia utilization in long haul journeys.  
Dedicated short range communication (DSRC) based on IEEE 802.11p was released in early 2000's as an initial support for V2X communication \cite{DSRC}. However, it was later shown in \cite{DSRC-1} that DSRC cannot guarantee the quality-of-service (QoS) requirements and it suffers from unbounded latency. Researchers, however advocate cellular solutions as preferable V2X standard \cite{cellular}, primarily because of the wide infrastructure deployment \cite{cellular-infra} and need for ultra-reliable and low-latency communication. 

5G is intended to support a diverse range of services, however initial solutions are expected to provide support for eMBB and URLLC use cases \cite{qualcomm,ngmn}. Provisioning of infotainment services for the passengers of autonomous vehicles falls under the eMBB use case of the 5G systems \cite{Metis}. The increasing popularity of video streaming services are forcing network operators and service providers to ensure a certain quality of experience (QoE) for infotainment activities. Due to the limited nature of wireless resources and challenging wireless environment, satisfying each users demand is a critical issue for which efficient radio resource allocation techniques play an important role \cite{radio}. In this regard, the existing literature on V2X communication focuses on several challenges including resource allocation, power minimization, ensuring reliability and network slicing \cite{petri1,aalborg,QoE,ikram,claudia,mehdi}.
Performance of cellular V2X communication is analyzed for different receiver types in \cite{petri}. An offloading mechanism for vehicles with low signal-to-interference ratio (SIR) is proposed in \cite{petri1}. QoE-aware power allocation scheme for device-to-device (D2D) video transmissions is studied in \cite{QoE}, with the goal of increasing user experience. Authors in \cite{lte-perf} evaluate the performance of LTE downlink unicast and multicast, based on single-cell point-to-multipoint (SC-PTM), and V2X sidelink transmission. A decentralized resource allocation mechanism to minimize network-wide power consumption while satisfying the constraints on queuing latency and reliability of vehicular user equipments is presented in \cite{ikram}. A brief overview of radio resource management techniques for eMBB use cases is studied in \cite{aalborg}. To address the stringent requirement of vehicular networks, authors in \cite{claudia} propose the design of customized network slices tailored to fulfill the demands of different use cases i.e. URLLC, eMBB. All of these works assume a fixed underlying architecture which is not elastic (cannot be scaled) and author in \cite{claudia} propose a flexible architecture without analyzing its performance. Thus, in this work we propose the joint analysis of an elastic network where resource allocation is aimed to satisfy the user QoE.

\begin{table*}[ht]
\caption{Summary of existing literature.}
\centering
\label{param}
\begin{tabular}{l l l l l l l}
\hline
\textbf{Reference}   & \textbf{Network scenario}  & \textbf{UE type} & \textbf{Service type} & \textbf{Objective} & \textbf{Procedure}             \\ \hline \hline
\cite{QoE} & Single-cell & Device-to-device (D2D) & Video & QoE enhancement & PC, RA \\
\cite{LR} & Single-cell & Mobile users & Video & Video adaptation  & PC, RA \\
\cite{LR1} & Multi-cell & Mobile users & Video & QoE \& EE  & PC, RA \\
\cite{LR2} & Single-cell & D2D unlicensed & Video + FD & QoE aware  &  RA \\
\cite{LR3} & Femto multi-cell & Mobile users & Video + HTTP + FTP & QoE driven  &  RA \\
\cite{LR4} & Single-cell & Mobile users & Video + VoIP + FD & QoE driven & PC, RA \\
\cite{LR5} & Single-cell & Mobile users & Video + VoIP + HTTP & QoE aware & RA \\
\cite{LR6} & Multi-cell NOMA & Mobile users & Web browsing & QoE aware & UA, PC, RA \\
\hline
\vspace{0.1pt}
\end{tabular} \\
{\raggedright *PC: Power control, RA: Resource allocation, UA: User association, HTTP: Hypertext transfer protocol, FTP: File transfer protocol, FD: File download, VoIP: Voice over IP, NOMA: Non-orthogonal multiple access. \par}
\end{table*}

QoE enhancements have been studied from various perspectives in the existing literature \cite{QoE},\,\cite{LR}-\cite{LR6} and it is summarized in Table \ref{param}. The works \cite{QoE},\,\cite{LR},\,\cite{LR2},\,\cite{LR4},\,\cite{LR5} study the QoE enhancement in single-cell downlink networks for services such as video, VoIP, FD, and HTTP, via power control and resource allocation. Furthermore, the QoE enhancement in multi-cell scenarios is studied in \cite{LR1},\,\cite{LR3},\,\cite{LR6} for multiple services (web browsing, HTTP, and FTP). In multi-cell scenarios, the optimized variable for QoE enhancement were user association, power control, and resource allocation. All of these works assume that radio links are established with the access points and the QoE is enhanced on the basis of already established links. In our work, the first task is to offload the vehicles from weak V2I links to high quality V2V links, which is performed using a clustering algorithm. Then the video quality selection and resource allocation tasks are optimized to enhance QoE. Moreover, the existing literature assumes that the network topology is fixed during the communication duration i.e., access points does not change. In contrast, we have a flexible architecture where the access points for V2V communication are changed depending upon the link quality.

Vehicular use cases cover a multitude of scenarios spanning from self driving cars, to multimedia utilization on an in-car infotainment system, to real time diagnostics, and remote driving. The verticals for 5G vehicular devices are still to be disclosed, while network slicing is ramping up. Network slicing is the concept of creating multiple sliced networks on a shared physical infrastructure \cite{ns-def}. Slicing is visualized in \cite{Metis} as a means of providing service to several use cases with independent QoS constraints. A network slice can span across all the network entities including the core network (CN) and radio access network (RAN) \cite{ngmn-slicing}. The impact of CN slicing can be seen on control plane functionalities i.e. mobility management, authentication and configurable user plane functionalities. Slicing the RAN is more challenging due to the shared and time varying nature of wireless resources and it affects the time/frequency usage of wireless resources. Also it is possible to have a network slice spanning across the network i.e. CN and RAN but it will lead to higher complexity \cite{claudia}. 

In this paper, we introduce a novel resource allocation and video quality management (selection) algorithm for vehicular network considering the reliability requirements of V2X communication. This work considers a sliced vehicular network in the downlink direction consisting of a set of video streaming vehicles with multiple video qualities. A network-wide video quality and resource maximization problem is considered in this work subject to probabilistic constraint on the received video duration. The constraint ensures that a vehicle must always buffer a certain amount of video frames to ensure seamless experience. In the proposed solution, the road side unit (RSU) is responsible for clustering the vehicles (with weak QoE) and allocating them to \emph{slice leaders} (SLs) which will relay the data to clustered vehicles. SLs are the vehicles with high QoE and good quality vehicle-to-vehicle (V2V) and vehicle-to-infrastructure (V2I) links making it suitable to serve as a relay. Clustered vehicles with weak QoE have the choice to either connect with the RSU or the SL and are categorized as \emph{free vehicle}. Vehicles which have high quality V2I link and can only connect to the RSU are termed as \emph{compelled vehicles} 

RSU is responsible for selecting the quality and resources of the vehicles connected to it directly while the decision for clustered vehicle is made by the SL. The formulated problem is solved using Lyapunov optimization which ensures that the constraints on queue length are satisfied \cite{neely}. Selection of resources and video quality is performed using the Lyapunov drift plus penalty at each time instant which satisfies the reliability requirements on queue length. The performance of the proposed algorithm is evaluated using a system level LTE-A compliant simulator built as per the guidelines of international telecommunication union (ITU) \cite{WINII}.

The rest of the paper is structured as follows. Section II explains the system model and presents the problem formulation. Section III discusses the network slicing and vehicle clustering mechanism in detail. Lyapunov optimization for vehicle scheduling and video selection is discussed in Section IV. Performance evaluation of the proposed resource allocation and video selection algorithm is provided in Section V and Section VI concludes the paper.

\begin{figure}[hbtp]
	\centering
	\includegraphics[width=0.5\textwidth]{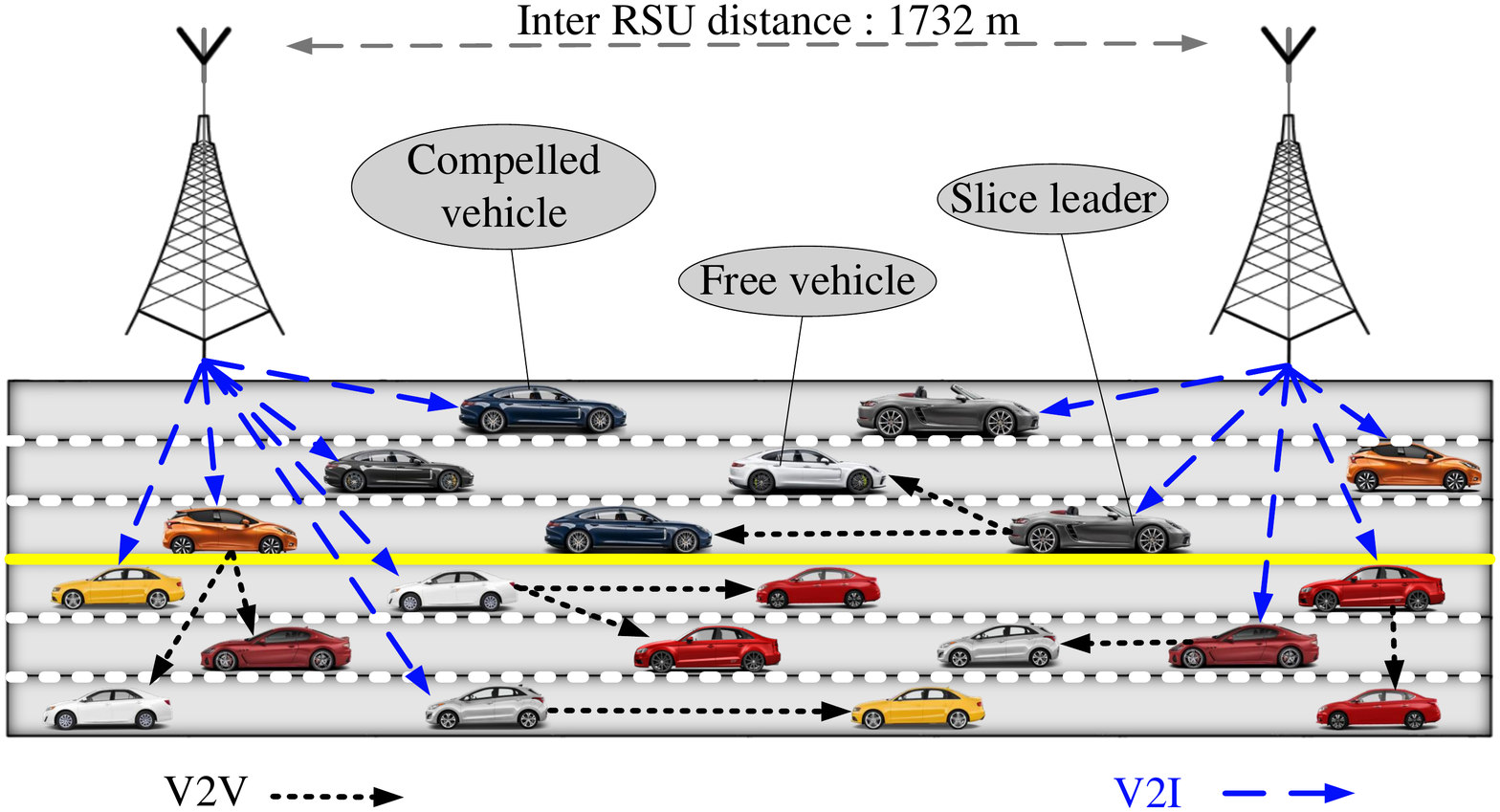}
	\vspace{-50pt}
	\caption{\label{Layout}Layout of the vehicular network.}
\end{figure}

\emph{Notations}: We will use boldface lower case letter \textbf{x} and boldface upper case letter \textbf{X} to represent vectors and matrices. The cardinality of set $\mathcal{X}$ is denoted as $X$.

%% file: system_model.tex
We study a downlink orthogonal frequency-division multiple access (OFDMA) system with single-input multiple-output (SIMO) transmission consisting of a set $\mathcal{B} = \{1,...,b,...,B\}$ of RSU and a set $\mathcal{V}$ of vehicles. Vehicles in the network can be categorized as \emph{slice leaders, free users} and \emph{compelled users}. SLs $\mathcal{S}$ have high quality V2I and V2V links they serve as virtual RSU for their neighboring vehicles. Free users are free to choose either RSU or SLs as their serving nodes as indicated using the link indicator variables $l_{bv}, l_{sv}$. The set of free users that are in the vicinity of SL $s$ is $\mathcal{F}_s$ and $\mathcal{F} = \{ \cup_{s\in\mathcal{S}} \mathcal{F}_s \}$ is the set of all free users in the network. Compelled users $\mathcal{C}$ have high quality V2I links and poor V2V links with their neighbors. RSUs and SLs have their own sets of resources $\mathcal{M}^{\text{RSU}}$ and $\mathcal{M}^{\text{SL}}$ which consists of equal number of physical resource blocks (PRB). 
\textcolor{black}
{
Therein, vehicle clustering is performed to improve the quality of links between vehicular users and their serving nodes yeilding enhanced QoE. First, the vehicles with weak V2I links are identified and nearby vehicles are grouped together on the basis of the neighborhood size parameter. Then, each cluster is assigned to a close-by vehicle with strong V2I and V2V links known as slice leader. This allows clustered vehicles to establish strong V2V links as opposed to their weak V2I links. Finally, the serving nodes of the clustered vehicle are changed from RSU to SL. Vehicles in the network are partitioned into three sets based on their operation mode: 
\begin{itemize}
  \item V2I downlink only: Compelled vehicle $\mathcal{C}$
  \item V2V downlink only: Free vehicle $\mathcal{F}$
  \item V2I downlink and V2V uplink: Slice leader $\mathcal{S}$
\end{itemize}
} 

\textcolor{black}
{
This partitioning is formally represented in (1a). Constraints (1b)-(1d) ensure that vehicles connect to one service point only i.e., slice leaders and compelled vehicles are served by the RSU and free vehicles are served by their corresponding slice leaders. Finally, the link indicator constraint (1e) ensures that only one link is used at any time, either V2I or V2V in which case the corresponding indicator has a value 1. 
}
\begin{subequations}
\label{sets}
\begin{align}
\label{set-partition} \mathcal{V} = {\mathcal{S} \cup \mathcal{F} \cup \mathcal{C}} \hspace{10pt} \text{with} &\hspace{10pt} |\mathcal{V}| = |\mathcal{S}| + |\mathcal{F}| + |\mathcal{C}| \\
\mathcal{S}_b \cap  \mathcal{S}_b' = \emptyset \hspace{10pt} & \forall   b, b' \in \mathcal{B}, b \neq b' \\
\mathcal{C}_b \cap  \mathcal{C}_b' = \emptyset \hspace{10pt} & \forall   b, b' \in \mathcal{B}, b \neq b' \\
\mathcal{F}_s \cap  \mathcal{F}_s' = \emptyset \hspace{10pt} & \forall  s, s' \in \mathcal{S}, s \neq s' \\
\label{link} l_{bv} + l_{sv} = 1 \hspace{10pt} &  \forall l_{bv}, l_{sv} \in \{ 0,1 \}  
\end{align}
\end{subequations}

\textcolor{black}
{
For such partitioning of vehicles, we use spectral clustering as explained in Algorithm 1. An important input parameter to the algorithm is the neighborhood size $\sigma$ which controls the similarity among vehicles with given locations. A lower value of $\sigma$ results into large number of clusters with smaller number of vehicles per cluster. On the other hand, large value of $\sigma$ results in few clusters with large number of vehicles. The importance of $\sigma$ can be highlighted in an abstract manner as follows:
}

\textcolor{black}
{Using small value of $\sigma$ in a sparse network can lead to clusters with only one vehicle, creating the same number of clusters as the number of nodes. While, using large value of $\sigma$ in a denser network increases the similarity among vehicles and could lead to a one cluster scenario. Moreover, when we have large number of clusters, the number of slice leader also increases which increases the interference in the network due to resource reuse. Therefore, the choice of the trade off parameter $\sigma$ directly impacts the performance of the proposed algorithm. }

The wireless channel between RSU and vehicles is modeled using geometry-based stochastic channel model (GSCM) given by \cite{WINII} and the path loss model follows the macro to relay model. Communication channel gain from RSU $b$ to vehicle $v \in \mathcal{V} $ on resource block $m \in \mathcal{M}^{\text{RSU}}$ is denoted as $\boldsymbol h_{bv}^m(t)$. On the other hand SLs and vehicles have independent and identically distributed (i.i.d.) channels and they follow the V2V path loss model given in \cite{V2V_pathloss}. The channel from SL $s$ to vehicle $v \in \mathcal{F}$ is denoted as $\boldsymbol h_{sv}^m(t)$, where $m \in \mathcal{M}^{\text{SL}}$ is a RB available for V2V communication. Assignment of physical resources for receiver vehicle $v$ is indicated using a resource utilization vector $\boldsymbol x_{v}(t) = [x_{v}^{m}(t)]_{m \in \mathcal{M}^{\text{RSU}}} $  with $x_{v}^{m} = 1$ when vehicle $v \in \mathcal{V} $ is assigned with resource $m$ and $x_{v}^{m} = 0$ otherwise. 
The instantaneous data rate of vehicle $v \in \mathcal{V}$ is:

\begin{align}
\begin{split}
r_v(x_v^m) & = \sum_{b \in \mathcal{B}} l_{bv} \bigg( \sum_{m \in \mathcal{M}^{\text{RSU}}} x_{v}^{m} \phi \log_{2} \big( 1 + \frac{p_{bv}^{m}|h_{bv}^{m}|^2}{\sigma^{2} + I_{bv}^{m}} \big) \bigg) \\ 
& + \sum_{s \in \mathcal{S}} l_{sv} \bigg(\sum_{m \in \mathcal{M}^{\text{SL}}} x_{v}^{m} \phi \log_{2} \big( 1 + \frac{p_{sv}^{m} |h_{sv}^{m}|^2}{\sigma^{2} + I_{sv}^{m}} \big)\bigg), \\
\end{split}
\end{align}
where $\phi$ is the bandwidth of a physical resource block. The transmission power assigned by RSU is $p_{bv}^{m}(t) \in [0,p_{b}^{max}]$ and the transmission power of slice leader is $p_{sv}^{m}(t) \in [0,p_{s}^{max}]$. Interference in the network occurs when the RBs are reused by different RSUs and SLs. $I_{bv}^{m}(t) = \sum_{\mathcal{B} \backslash \{b\}} \sum_{v' \in \mathcal{V} \backslash \{v\}} x_{v'}^{m}(t) p_{b'v'}^{m}(t) |h_{b'v}^{m}(t)|^2$, $b' \in \mathcal{B} \backslash \{b\}$ is the interference at vehicle $v$ over the resource block $m$ when served by RSU $b$. Similarly, the interference experienced by vehicle $v$ served by SL $s$ is denoted as $I_{sv}^{m}(t) = \sum_{\mathcal{S} \backslash \{s\}}\sum_{v' \in \mathcal{F} \backslash \{v\}} x_{v'}^{m}(t) p_{s'v'}^{m}(t) |h_{s'v}^{m}(t)|^2$.

\begin{figure}[hbtp]
	\centering
	\includegraphics[width=0.45\textwidth]{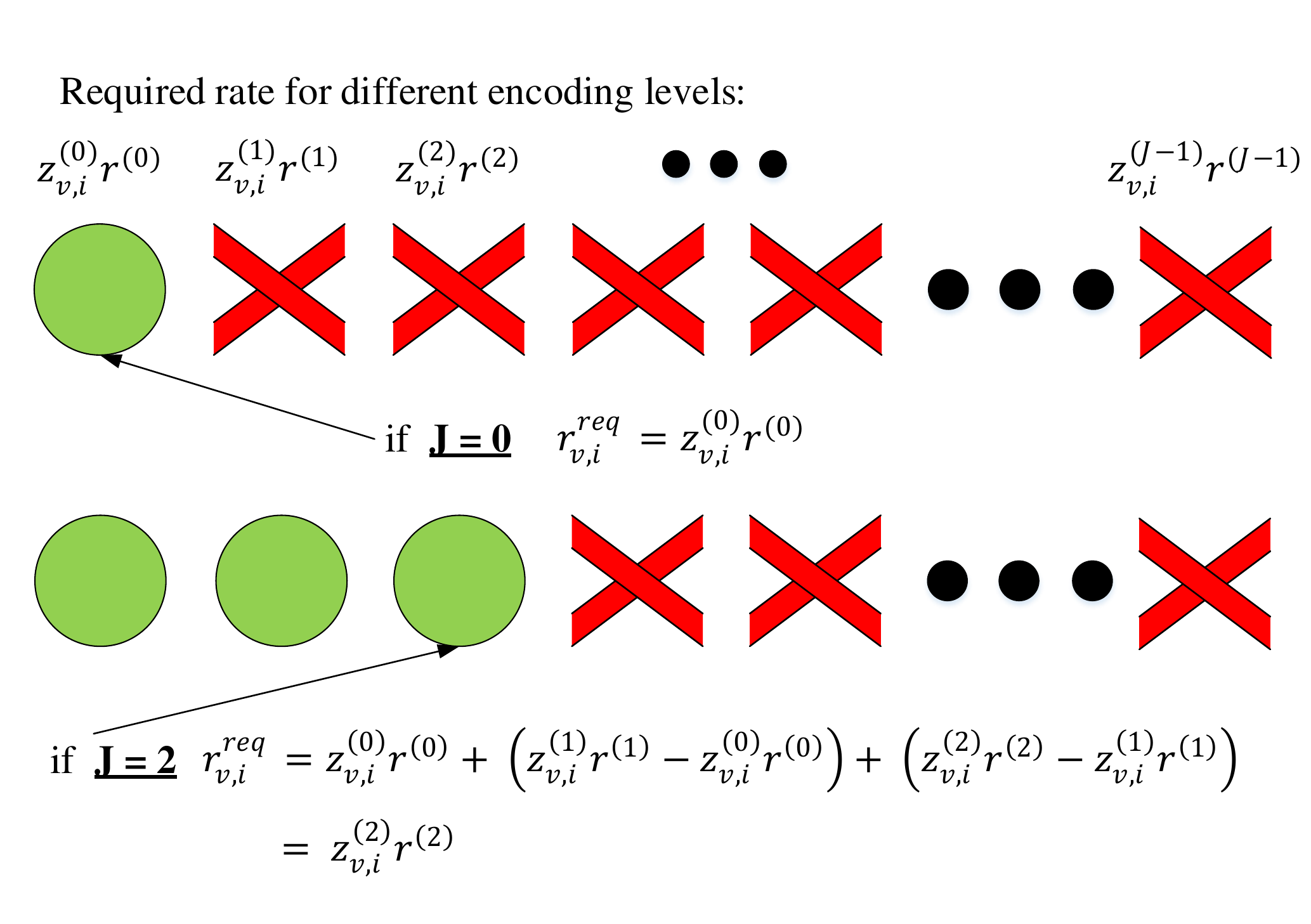}
	\caption{\label{req_rate}Choice of required rate for the encoded video quality.}
\end{figure}

In the past few years much of the attention is paid to the development of chunk based video encoding schemes, where a video is first divided into chunks and then every chunk is encoded into multiple quality levels. Therefore, the video streaming logic that resides either at the client or the server needs to make a fine grained decision of the quality per chunk. In our formulation, we first assume that the video is divided into $I$ chunks. Moreover, we assume that each video chunk is encoded into J quality levels with rates $r_0,r_1,r_2,....,r_{J-1}$, and we denote the set of encoding levels by $\mathcal{J} \in \{0,1,2,...,J-1\}$. A decision variable $z_{v,i}^{(j)}  \in \{ 0,1 \}$ is used to indicate that vehicle $v$ is fetching video chunk $i$ at quality level $j$. Moreover, each vehicle can only stream a video chunk with one quality, \textcolor{black}{i.e., the required rate to download chunk $i$ by vehicle $v$ is given as follows:} 
\textcolor{black}{
\begin{equation}
\label{eq_req_rate}
r_{v,i}^{\text{req}} = z_{v,i}^0 r^{(0)} + \sum_{j=1}^{J-1} \bigg( z_{v,i}^{(j)} r^{(j)} - z_{v,i}^{(j-1)} r^{(j-1)} \bigg)
\end{equation}
where, $z_{v,i}^0 r^{(0)}$ is the required rate for lowest video quality as shown in Fig. \ref{req_rate} and the required rate for highest video quality is $z_{v,i}^{(J-1)} r^{(J-1)}$. The value of decision variable $z_{v,i}^{(k)} = 1$ for all $k \leq j$ where $j$ is the selected chunk quality. The motivation behind \eqref{eq_req_rate} is that the vehicle remains a candidate to all the encoding levels that are below the selected quality. }

The network model assumes that each vehicle has queues of video frames maintained at their serving RSU and the arrival rate of video frames in each queue is equal to the rate required by the selected video chunk. The evolution of the queue length of $v$-\text{th} vehicle at the RSU $b$ is:
\begin{align}
q_{bv}(t+1) = [q_{bv}(t) - r_{bv}(t)]^{+} + r_{v,i}^{\text{req}}  & \quad v \in \mathcal{V},
\end{align}
where $r_{bv}(t)$ is the service rate of RSU $b$. 
Free vehicles which receive their video stream from a slice leaders have queues evolving at RSU and at their serving SL $s \in \mathcal{S}$ given as:
\begin{subequations}
\begin{align}
q_{bv}(t+1) = [q_{bv}(t) - r_{bs}^{v}(t)]^{+} + r_{v,i}^{\text{req}} & \quad v \in \mathcal{F}, \\
q_{sv}(t+1) = [q_{sv}(t) - r_{sv}(t)]^{+} + r_{bs}^{v}(t) & \quad v \in \mathcal{F},
\end{align}
\end{subequations}
where $r_{sv}(t)$ is the service rate with which SL relays the video frames to vehicle $v$ and $r_{bs}^{v}(t)$ represents the backhaul link between the SL and RSU. 
To ensure QoE at each vehicles, it is necessary to ensure queue stability at corresponding transmitters i.e. $\bar{q}_{bv} = \lim_{t\to\infty} \frac{1}{t} \sum_{\tau = 0}^{t-1} q_{bv}(\tau) \leq \infty$. On the other hand, queue stability for free vehicles $v \in \mathcal{F}$ is ensured when the queues at both RSU and SL are stable.

Vehicles which serve as SLs establish backhaul links with the RSU that should support the data rates of connected free vehicles along their own rate demands, i.e., 
\begin{align}
\qquad r_{bs} - \sum_{f \in \mathcal{F}_s} r_{sf}  \geq r_{s}^{\text{req}}  \quad \forall s \in \mathcal{S},
\end{align}

To ensure seamless video streaming each receiver must at least buffer video frames with playback time of $\psi$ in advance. Hence, we impose a probabilistic constraint on each vehicle  which ensures that playback video in its buffer is less than $\psi$ with a very small probability $\epsilon \ll 1$, 
\begin{align}
\label{probabilistic}
\qquad  \text{Pr}\bigg(\frac{\sum_{\tau=1}^{t}r_v(\tau)}{r_{v,i}^{\text{req}}} - t \leq \Psi\bigg) \leq \epsilon \quad \forall v \in \mathcal{V},
\end{align}
where $\frac{\sum_{\tau=1}^{t}r_v(\tau)}{r_{v,i}^{\text{req}}}$ is the duration of video frames transmitted to vehicle $v$ till elapsed time $t$. 

Each vehicle in the network can utilize the maximum number of resources available at their serving node. In this regard, \textcolor{black}{the per vehicle resource allocation constraints are defined as follows: 
\begin{subequations}
\label{resources}
\begin{align}
x_v^m \in \{0,1\}  \hspace{10pt} & \forall { m \in \{ M^{\text{RSU}}, M^{\text{SL}}\}}, \\
\textstyle{\sum_{m}} x_v^{m} \textstyle{\sum_{m'}} x_v^{m'} = 0 \hspace{10pt} & \forall m \in \mathcal{M}^{\text{RSU}}, m' \in \mathcal{M}^{\text{SL}}, \\
\textstyle{\sum_{m}} x_v^{m} = M^{\text{RSU}} \hspace{10pt} & \forall m \in \mathcal{M}^{\text{RSU}}, \\
\textstyle{\sum_{m}} x_v^{m} = M^{\text{SL}} \hspace{10pt} & \forall m \in \mathcal{M}^{\text{SL}}, 
\end{align}
\end{subequations}
where \eqref{resources} implies that resources of the RSU and SL are not shared by any vehicle i.e., the SL serves only those free vehicles which are connected to it and the rest are served by the RSU.} 

Furthermore, each resource of RSU and SL can be only used by one vehicle at each time instant, i.e.,
\begin{equation}
\begin{aligned} 
  \sum_{v \in \mathcal{V}\backslash\mathcal{F}} & x_v^{m} \leq 1 \qquad m \in M^{\text{RSU}}, \\
 \sum_{v \in \mathcal{F}} & x_v^{m} \leq 1 \quad \quad m \in M^{\text{SL}},
\end{aligned}
\end{equation} 
Note that the constraints (8) and (9) ensure that the constraint of link indicators (1e) is satisfied. 

\begin{figure}[hbtp]
	\centering
	\vspace{-10pt}
	\includegraphics[width=0.5\textwidth]{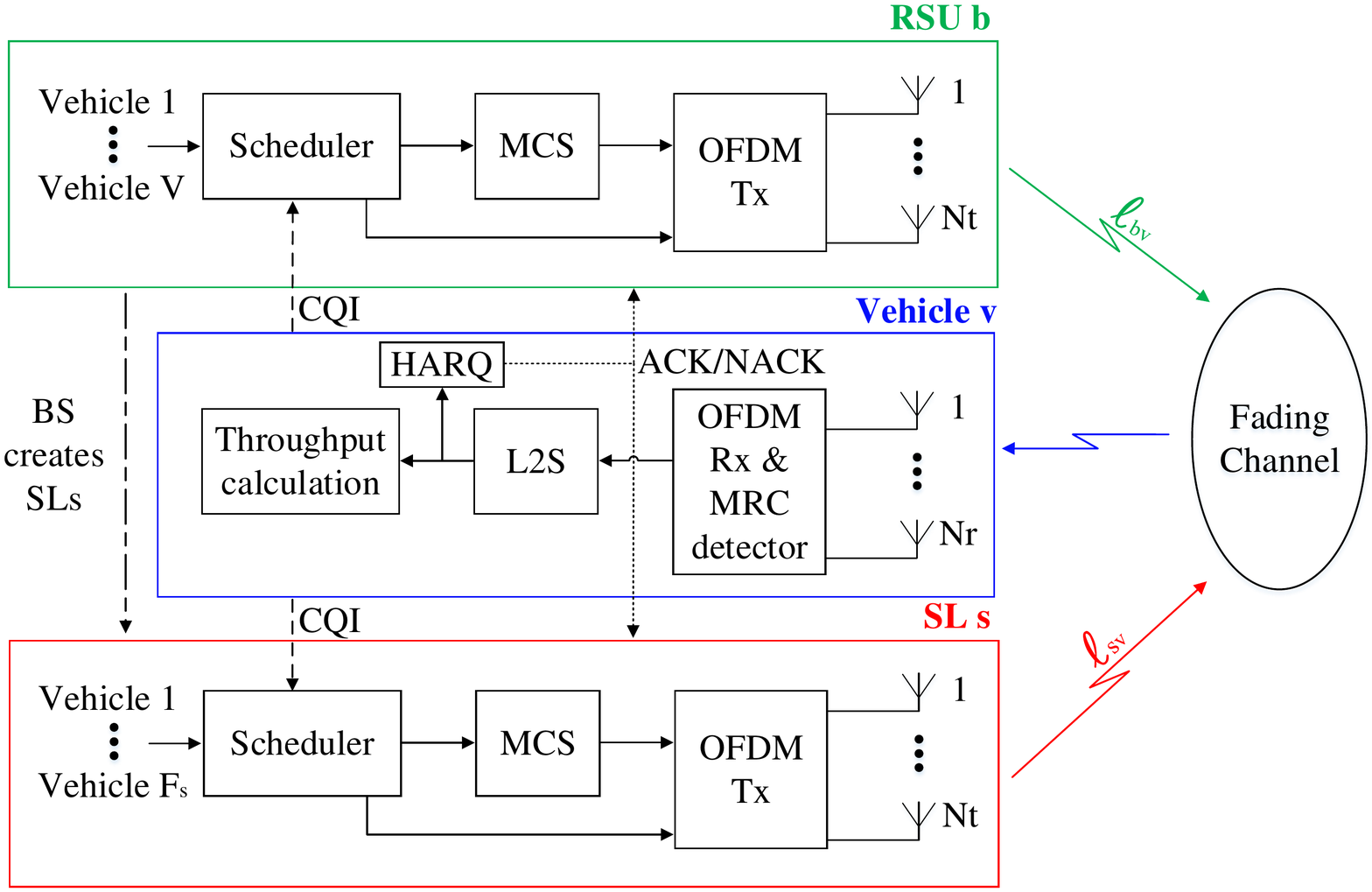}
	\vspace{-20pt}
	\caption{\label{Link_Model}Link model of the vehicular network with network slicing.}
\end{figure}

Fig. \ref{Link_Model} illustrates the link model between the RSU, SL and a vehicle. In the simulator link-to-system (L2S) interface is used for the exact modeling of radio links. An important input parameter for L2S interface is the channel quality information (CQI) which is transmitted by each vehicle and is used to decide the modulation and coding scheme (MCS). Intersymbol interference is mitigated by using cyclic prefix longer than the delay spread. To exploit spatial diversity the vehicles use maximal ratio combining (MRC) receivers. Calculation of SINR is performed for each RB. The symbols are perfectly synchronized in time and frequency domain. Computational overhead is minimized by using mutual information based effective SINR mapping (MIESM) which maps the SINR to respective mutual information curve. Erroneous transmissions are detected by mapping the frame error curve to the corresponding mutual information curve. Successful transmission of video frames generates an acknowledgment (ACK) while failed transmissions are represented with negative acknowledgement (NACK). Re-transmission of the failed video frames is handled by the hybrid automatic repeat request (HARQ).

\subsection{Problem Formulation}
The design criteria is to maximize the quality of experience (QoE) for each vehicle. QoE in this work is defined as a weighted sum of two terms, the first term takes care of maximizing the average quality while preserving fairness among users $\sum_j \gamma^j z_{v,i}^j$, where $\gamma \ll 1$ \cite{anis,anis-1}. The second one, penalize the quality switching between adjacent chunks and it is captured by: $-\beta \mathbbm{1} (Z_{v,i} <  Z_{v,i-1})$, where $Z_{v,i} = j$ if $z_{v,i}^j = 1$, and 0 otherwise, where $0 \leq \beta \leq 1$. Therefore, for every chunk $i$ to download, we solve the following optimization problem:

\textcolor{black}{
\begin{equation}
\begin{aligned} 
\label{objective-function}
\max_{\substack{ x_{v}^{m}, z_{v,i}^{(j)} \\ \mathcal{C}, \hspace{1pt} \mathcal{S}, \hspace{1pt} \mathcal{F}}} & \sum_v \left \{ \sum_{j}  { (\gamma^j z_{v,i}^j) - \beta  \bar{g}(Z_{v,i}, Z_{v,i-1}) }\right \}, \\
 & \text{subject to} \qquad \text{(1)}, \text{(3)}, \text{(6)} , \text{(7)}, \text{(8)}, \text{(9)}
\end{aligned}
\end{equation}}

The value of indicator function is 1 whenever the quality decreases compared to the quality at previous video chunk otherwise it is 0. Here, $\beta$ is the tradeoff parameter that controls the impact of video fluctuation on the objective. 

The indicator function in our optimization problem is transformed into an equivalent s-curve function \cite{sigmoid} i.e.,
\begin{align}
 g(Z_{v,i}, Z_{v,i-1}) = \frac{e^{-\alpha({Z_{v,i} - Z_{v,i-1}})}}
 {1+e^{-\alpha({Z_{v,i} - Z_{v,i-1}})}},
\end{align}
where $\alpha$ controls the slope of the s-curve. To solve the objective function in (\ref{objective-function}) the slicing algorithm determines the partition of sets $\mathcal{S} ,\mathcal{F} ,\mathcal{C}$. When the set partitions are determined RSU solves the network wide optimization problem and performs resource allocation and selection of video quality while ensuring queue stability.

%% file: vehicle_clustering.tex
Network slicing is defined as a virtualization capability that slices the physical network into several logical networks which can be independently optimized. Network slicing brings elasticity which will be helpful in addressing the issues of efficiency, flexibility and scalability for the future generation of mobile services. The key enabler of this technology is network functionality virtualization (NFV) which can reconfigure networks with software and manage network resources. In the present network architecture we have predefined service function chains for individual services. 

Network slicing enables the reconfiguration of service functions via NFV, with which the same infrastructure can be used to provide different services. The key step in network slicing is to determine clusters of free vehicles. RSU measures the CQI value of all the vehicles and computes the similarity between the low SINR vehicles (with weak QoE) in the network based on the geographical information. RSU constructs the distance based similarity matrix \textbf{C} in a way that vehicles which have large distance have less similarity and vice versa. Since the objective function in (\ref{objective-function}) requires the partition of set $\mathcal{V}$ as per (\ref{sets}), we decouple  (\ref{objective-function}) and first utilize spectral clustering algorithm to determine $\mathcal{F}$ \cite{spectral-clustering}.

\begin{algorithm}
 \caption{Partitioning Free Users}
 \begin{algorithmic}[1]
 \renewcommand{\algorithmicrequire}{\textbf{Input:}}
 \renewcommand{\algorithmicensure}{\textbf{Output:}}
 \REQUIRE Similarity matrix \textbf{C}
 \ENSURE  a set of k clusters i.e. $\cup \mathcal{F}_k$
 \\ \textit{Initialization}:
  \STATE Let W be the weighted adjacency matrix of \textbf{C}.
  \STATE Compute the unnormalized Laplacian i.e \textbf{L} = \textbf{D} - \textbf{W}, where \textbf{D} is a degree matrix of \textbf{C}.
  \STATE The number of clusters k corresponds to the index of maximum eigenvalue of \textbf{L}.
  \STATE Let \textbf{U} $\in \mathbb{R}^{n\text{x}k} $ represent the k eigenvectors of \textbf{L} i.e. $u_1, ... , u_k$ corresponds to the columns of \textbf{U}.
  \FOR {$i = 1,...,n$}
  \STATE let $x_i \in \mathbb{R}^k$ be the vector representing the $i$-th row of \textbf{U}.
  \STATE Cluster the points $(x_i)_{i=1,...,n}$ in $\mathbb{R}^k$ with the k-means algorithm into clusters $\mathcal{F}_1,..., \mathcal{F}_k$.
  \ENDFOR
 \RETURN $\mathcal{F} = \{\mathcal{F}_1,...,\mathcal{F}_k \}$ 
 \end{algorithmic} 
 \end{algorithm}

Euclidean distance based similarity matrix $\textbf{C}$ is formulated at each RSU, where $d_{vv'}$ corresponds to the $(v,v')$-th entry. Similarity \textbf{C} two vehicles $v,v' \in \mathcal{V}$ is measured by the Gaussian similarity function given as \cite{gaussian}: 

\begin{equation}
d_{vv'} = \exp \bigg( \frac{-||{d}_{v} - {d}_{v'}||}{2\sigma^{2}} \bigg)
\label{receivedsignal},
\end{equation}
where $d_v$ corresponds to the location of vehicle $v$. Impact of neighborhood size is controlled by $\sigma$. Small value of $\sigma$ results into less vehicles per cluster and vice versa. Apart from $\sigma$, an important clustering parameter is the input $k$ which determines the number of clusters. If $\sigma$ is fixed, an appropriate choice of clusters $k$ is determined by eigenvalue method. Spectral clustering exploits the geometry of nodes in a graph. The graph Laplacian matrix is formulated as \textbf{L} = \textbf{D} $-$ \textbf{S}, where \textbf{D} is diagonal matrix with $v$-th diagonal element given as $\sum_{v'=1}^{V} d_{vv'}$. For cluster $\mathcal{F}$, the combination of the smallest $\mathcal{F}$ eigenvectors of \textbf{L} can be used to determine the input parameter $k$ for $k$-means clustering. 

\begin{equation}\label{similarity}
k = \arg\max_{i}(\chi_{i+1}-\chi_i), \hspace{0.2cm} i = 1,...,n-1,
\end{equation}
where $\chi_i$ is the $i$-th smallest eigenvalue. When the nodes on the graph are uniformly distributed as $k$ clusters, the first $k$ eigenvalues are small and the ($k+1$)th eigenvalue becomes relatively large. Since the spectral clustering algorithm requires the knowledge of similarity matrix \textbf{S}, it is categorized as a centralized clustering mechanism. The output of clustering algorithm is the set of free users $\mathcal{F} = \{ \cup \mathcal{F}_k \}$.

\begin{algorithm}
 \caption{Slicing Algorithm}
 \begin{algorithmic}[1]
 \renewcommand{\algorithmicrequire}{\textbf{Input:}}
 \renewcommand{\algorithmicensure}{\textbf{Output:}}
 \REQUIRE {$\mathcal{F}$}
 \ENSURE  $\mathcal{S}$
 \\ \textit{Initialization}:
  \FOR {each cluster $i = 1,...,k$}
  \STATE Let $y_i$ represent the center of cluster $i$
  \STATE Find the distance between $y_i$ and vehicles $v \in \mathcal{V} \backslash \mathcal{F}$
  \STATE Find $s_i = \text{arg min}_{\forall v \in \mathcal{V} \backslash \mathcal{F}}\{y_i - d_v\}$, where $d_v$ is the location of vehicle $v$
  \ENDFOR
 \RETURN $\mathcal{S} = \{s_1,...,s_k\}$ 
 \end{algorithmic} 
 \end{algorithm}

To serve the set of free vehicles we need to determine the slice leaders $\mathcal{S}$. As described earlier $\mathcal{S}$ is the set of vehicle which have high quality V2I link and V2V link, which makes it suitable for them to serve as an access point. RSU utilizes the geographical information to determine the distance of all vehicles $v\in \mathcal{V} \backslash \mathcal{F} $ from cluster center. Since the vehicular devices are mobile, it is highly likely that the nearest vehicle from cluster center will have high quality V2V links with all the vehicles of the respective cluster. A vehicle which is closest to the cluster center then becomes the slice leader as per the slicing algorithm. After determining the free users $\mathcal{F}$ and slice leaders $\mathcal{S}$ RSU utilizes (\ref{sets}) to find the compelled vehicles $\mathcal{C}$. In the upcoming section Lyapunov optimization to jointly optimize resources and video quality is performed.

%% file: lyapunov.tex
The Lyapunov drift method is widely used to solve optimization problems that evolve over time. Since the problem in the current work has time varying channel states, we leverage Lyapunov drift framework to ensure queue stability and performance optimization. In the Lyapunov method the choice of the optimal value of penalty comes with a trade off between the network stability and average queue backlog. For tractability of the inequality constraint (\ref{probabilistic}) we resort to the Markovian inequality. Markovian inequality specifies that for a non-negative random variable X  and $a \geq 0$ we have \text{Pr}$(\text{X} \geq a) \leq \mathbb{E}[X]/a$ \cite{markov}. Applying the Markovian inequality on (\ref{probabilistic}) yields, 

\begin{align}
\text{Pr} \hspace{1pt}(q_{bv}(T) \geq q_{bv}(0) - \Psi r_{v,i}^{\text{req}}) \leq \frac{\mathbb{E} [q_{bv}(T)]}{(q_{bv}(0) - \Psi r_{v,i}^{\text{req}})} \leq \epsilon
\end{align}

Note that the stability of virtual queues ensures the stability of the objective function and the satisfaction of probabilistic constraints. Two sets of virtual queues $U_v$ and $Y_v$ for vehicle $v \in \mathcal{V}$ connected to RSU and for vehicle $v \in \mathcal{F}$ connected to SL are introduced. The evolution of virtual queues is given as:

\begin{align}
\label{virtualq}
\notag U_{v}(T+1) = & [U_v(t) + q_{bv}(t+1) - \epsilon (q_{bv}(0) - \Psi r_{v,i}^{\text{req}})] ^{+},\\ 
\notag Y_{v}(T+1) = & [Y_v(t) + q_{bv}(t+1) + q_{sv}(t+1) \\
& - \epsilon (q_{bv}(0) +q_{sv}(0) - \Psi r_{v,i}^{\text{req}})]^{+},
\end{align}

Furthermore, as per (\ref{virtualq}), the stability of virtual queues ensure the stability of actual queues, in which case the stability of queues given in (\ref{virtualq}) is sufficient for the stability of network. 

We define a collective queue vector $\Theta(t) \overset{\Delta}{=} [U_v(t), Y_v(t), \boldsymbol x_v^{av}, \boldsymbol z_{v,i}^{av} ]$ including the current running time averages of control vectors $\boldsymbol x_v, \boldsymbol z_{v,i}^{(j)}$ defined as:
\begin{subequations}
\begin{align}
\boldsymbol x_{v}^{av}(t) \overset{\Delta}{=} \frac{1}{t} \sum_{\tau = 0}^{t-1}x_{v}^{m}(\tau) \qquad v \in  \mathcal{V},\\
\boldsymbol z_{v,i}^{av}(t) \overset{\Delta}{=} \frac{1}{t} \sum_{\tau = 0}^{t-1}z_{v,i}^{(j)}(\tau) \qquad v \in  \mathcal{V},
\end{align}
\end{subequations}

The collective queue vector represents the multi-dimensional state of the system, which is composed of the virtual queues (15), and the time-averages of control variables (16). The non-negative Lyapunov function for the collective queue vector is \cite{neely}:
\begin{align}
L(\Theta(t)) \overset{\Delta}{=} \frac{1}{2} \bigg (\sum_{v \in \mathcal{V} \backslash \mathcal{F}} U_v(t)^2 + \sum_{v \in \mathcal{F}} Y_v(t)^2 \bigg )
\end{align}

The conditional Lyapunov drift $\Delta (\Theta(t))$ is \cite{neely}:
\begin{align}
\begin{split}
\label{drift}
& \Delta(\Theta(t))  \leq  \varphi - \mathbb{E} \{ \sum_{v \in \mathcal{V} \backslash \mathcal{F}} r_v[U_v(t) + q_{bv}(t) - \epsilon q_{bv}(0)]\\ 
& - r_{v,i}^{\text{req}}[U_v(t) + q_{bv}(t)  + \epsilon \psi(U_v(t) + q_{bv}(t) - \epsilon q_{bv}(0))  \\
& -\epsilon q_{bv}(0) ] + r_v[r_{v,i}^{\text{req}} ( 1 + \epsilon \psi  )] \} - \mathbb{E} \{ \sum_{v \in \mathcal{F}} r_v[r_{v,i}^{\text{req}} ( 1 + \epsilon \psi )] \\
&  + r_v[Y_v(t) + q_{bv}(t) + q_{sv}(t) - \epsilon(q_{bv}(0) + q_{sv}(0))]\\
& - r_{v,i}^{\text{req}}[Y_v(t) + q_{bv}(t) + q_{sv}(t) -\epsilon (q_{bv}(0)+q_{sv}(0))  \\
&   + \epsilon \psi(Y_v(t) + q_{bv}(t) + q_{sv}(t) - \epsilon (q_{bv}(0) + q_{sv}(0)) ]  \},  
\end{split}
\end{align}
where $\varphi$ is a constant which bounds the square terms of queues and the optimization variable i.e. $\{q_{bv}(t), q_{sv}(t), Y_v(t), U_v(t), r_{v,i}^{\text{req}}, r_v\}$. 

Introducing the penalty term with the objective function and minimizing the upper bound on Lyapunov drift plus penalty at each time $t$ yield the solution for the  optimization problem described in (\ref{objective-function}). 

\begin{equation}
\begin{aligned}
\label{avg-obj}
& \min_{x_{v}^{m},z_{v,i}^{(j)}} \{ -\eta[ \frac{\partial f(\boldsymbol x_v^{av}, \boldsymbol z_{v,i}^{av})}{\partial x_{v}^{m}} + \frac{\partial f(\boldsymbol x_v^{av}, \boldsymbol z_{v,i}^{av})}{\partial z_{v,i}^{(j)}}] + \Delta(\Theta(t)) \},  
\end{aligned}
\end{equation}
where $f(\boldsymbol x_v^{av}, \boldsymbol z_{v,i}^{av})$ is the objective function defined in (\ref{objective-function}) for the current running time averages and $\eta$ is a non-negative penalty parameter that controls the tradeoff between the optimality of the solution and average queue congestion. The optimization problem in (\ref{avg-obj}) requires the minimization of the upper bound on Lyapunov drift. Since minimization requires the objective function to be convex and in our problem the non-convexity in the Lyapunov drift expression is due to the product of optimization variables $x_{v}^{m}$ and $z_{v,i}^{(j)}$ shown as, 

\begin{align}
\label{non-convex-only}
r_v r_{v,i}^{\text{req}}(1+\epsilon\psi) = \sum_{m,j} x_{v}^{m} z_{v,i}^{(j)} \phi \log(1+\frac{p_{bv}^{m}|h_{bv}^{m}|^2}{\sigma^{2} + I_{bv}^{m}}) r^{(j)}  (1+\epsilon\psi)
\end{align}

\begin{figure*}
$\vartheta_{v}^{m} = \begin{cases}
 - \phi \log_{2} \big( 1 + \frac{p_{bv}^{m}|h_{bv}^{m}|^2}{\sigma^{2} + I_{bv}^{m}}\big) [U_v(t) + q_{bv}(t) - \epsilon q_{bv}(0)], \quad v \in \mathcal{V} \backslash \mathcal{F} \\
 - \phi \log_{2} \big( 1 + \frac{p_{bv}^{m}|h_{bv}^{m}|^2}{\sigma^{2} + I_{bv}^{m}}\big) [Y_v(t) + q_{bv}(t)  + q_{sv}(t)  - \epsilon(q_{bv}(0) + q_{sv}(0))], \hspace{9pt} v \in \mathcal{F}
\end{cases}
$\\
$\Phi_{v}^{j} = \begin{cases}
  -\eta \sum_j \gamma^{(j)} z_{v,i}^{av} +  \sum_{j} r^{(j)} [U_v(t) + q_{bv}(t)  + \epsilon \psi(U_v(t) + q_{bv}(t) - \epsilon q_{bv}(0)) -\epsilon q_{bv}(0)],  \quad v \in \mathcal{V} \backslash \mathcal{F}; \\
  -\eta \sum_j \gamma^{(j)} z_{v,i}^{av} +  \sum_{j} r^{(j)} [Y_v(t) + q_{bv}(t) + q_{sv}(t) + \epsilon \psi(Y_v(t) + q_{bv}(t) + q_{sv}(t) - \epsilon (q_{bv}(0) + q_{sv}(0)) \\
  -\epsilon (q_{bv}(0)+q_{sv}(0))],  \qquad  v \in \mathcal{F}; \\
\end{cases}$
\end{figure*}

Let $\zeta_{bv}^{m(j)} =\phi \log(1+\frac{p_{bv}^{m}|h_{bv}^{m}|^2}{\sigma^{2} + I_{bv}^{m}}) r^{(j)}(1+\epsilon \psi)$ and applying the equality $-4xz =  (x - z)^2 - (x + z)^2$ to (\ref{non-convex-only}) for simplifying various relations,
\begin{align}
\label{equality-non-convex}
\sum_{m,j} x_{v}^{m} z_{v,i}^{(j)} \zeta_{bv}^{m(j)} = \sum_{m,j} (\frac{\zeta_{bv}^{m(j)}}{4} )[( x_{v}^{m} - z_{v,i}^{(j)})^2 - ( x_{v}^{m} + z_{v,i}^{(j)})^2 ]
\end{align}

By applying the first order taylor expansion around ($a,b$) on (\ref{equality-non-convex}) we can approximate the non-convexity in (\ref{drift}) as,

\begin{equation}
\begin{aligned}
\label{taylor-non-convex}
\Gamma_{v}^{m(j)} & = \sum_{m,j}  (\frac{\zeta_{bv}^{m(j)}}{4}) \big[ (a_{v}^{m} + b_{v,i}^{(j)})^2 + 2(x_{v}^{m} - a_{v}^{m})(a_{v}^{m} + b_{v,i}^{(j)}) \\
 & + 2(z_{v,i}^{(j)} - b_{v,i}^{(j)})(a_{v}^{m} + b_{v,i}^{(j)})  -( x_{v}^{m} - z_{v,i}^{(j)})^2 \big]
\end{aligned}
\end{equation}

Concave-convex procedure (CCP) is used to find the optimal solution of upper bound minimization problem. The linear approximation in (\ref{taylor-non-convex}) is bounded by an additional variable $\rho_{v}^{m(j)} \geq \Gamma_{v}^{m(j)} $ which is minimized in the objective function. The proposed algorithm takes control actions at every time slot in reaction to the previous observations and queue states, to minimize the drift plus penalty expression written in terms of the optimization variables as:

\begin{equation}
\begin{aligned}
& \min_{x_{v}^{m},z_{v,i}^{(j)}} \sum_{v \in \mathcal{V}} x_{v}^{m} \vartheta_{v}^{m} + z_{v,i}^{(j)} \Phi_{v}^{(i)} - \rho_{v}^{m(j)} , \\
 & \text{subject to} \qquad \text{(1)},\text{(3)},\text{(6)} - \text{(9)}
\end{aligned}
\end{equation}

where $\vartheta_{v}^{m},\Phi_{v}^{j}$ represents the coefficients of $ x_{v}^{m}$ and $z_{v,i}^{(j)}$ respectively and  $\rho_{v}^{m(j)}$ represents the CCP part. Next we analyze the performance of the proposed method for different vehicular densities, resources, neighborhood size and reliability ($\epsilon$).

%% file: performance_analysis.tex
In this section, we analyze the performance of network slicing based vehicular network simulated on system level simulator. The simulation considers the layout of a six lane highway as proposed in \cite{petri1}, where vehicles in three lanes are moving in left direction and the vehicle in remaining lanes are moving towards right with the speed of 140 km/h. RSU network is placed alongside the highway at a distance of 35 m and the inter RSU distance is 1732 m. 

The simulated layout assumes the highway is stretched over a distance of 10 km. To model different vehicular densities in the network we change the inter vehicular distance i.e. large inter vehicular distance results in sparse network and vice versa. To avoid the overhead of computing real time information of V2V links at each time slot RSU repeats the slicing procedure after every 100 ms. After updating the link information RSU runs the slicing algorithm to partition the vehicles into their respective sets. RSU uses transmit power of 46 dBm while vehicles use transmit power of 20 dBm. The rest of simulation settings are tabulated in Table \ref{params}. 

\begin{figure}[hbtp!]
	\centering
	\includegraphics[width=0.5\textwidth]{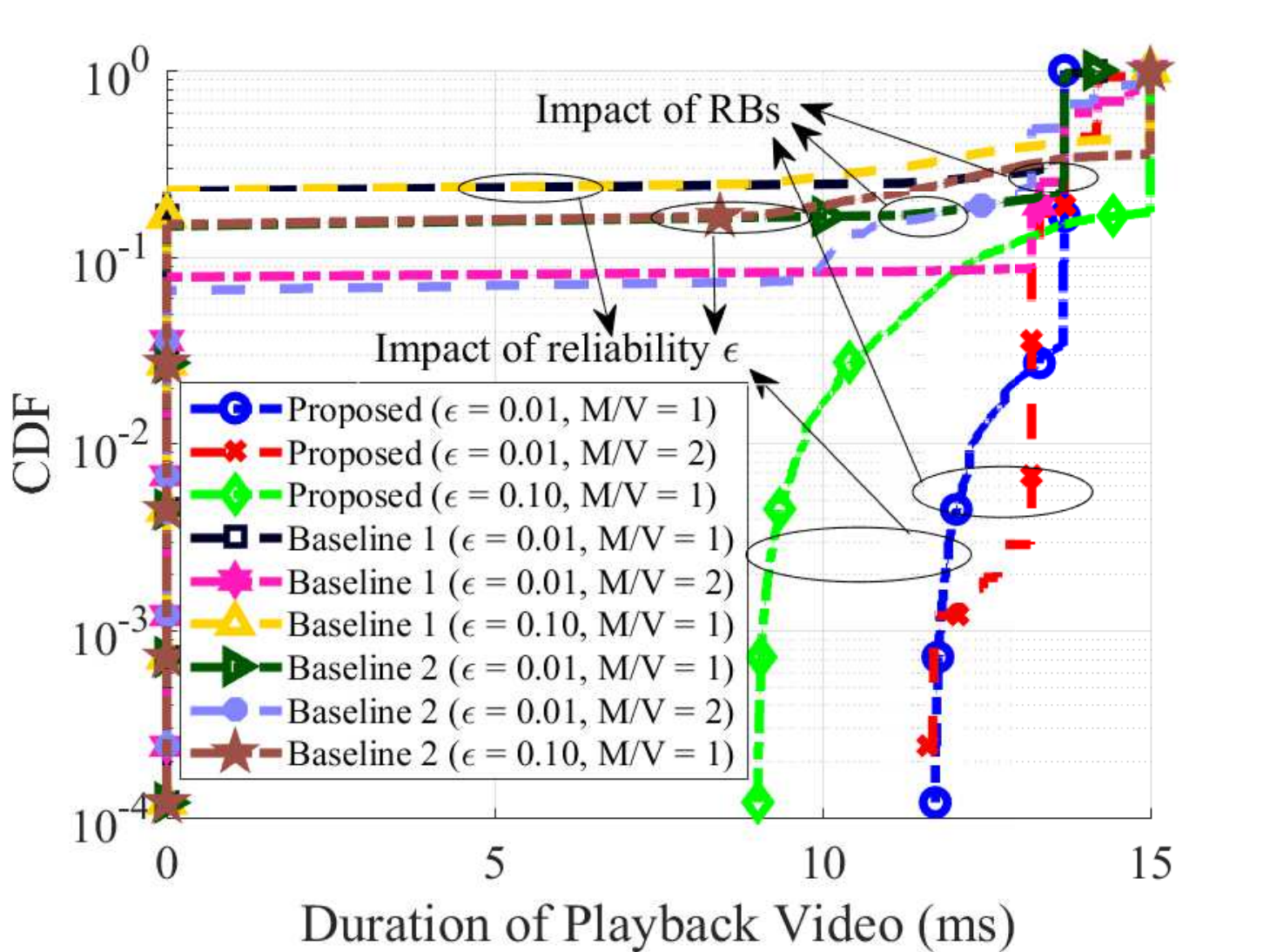}
	\caption{\label{rx_buff}CDF of playback video duration available at the receiver.}
\end{figure}

\textcolor{black}
{
The performance of the proposed network slicing solution is compared with baseline 1 and baseline 2. Baseline 1 from \cite{petri} was used to develop the vehicular communication in LTE-A system level simulator. Baseline 2 is from  \cite{petri1}, which  follows the concept of relaying vehicles to enhance cell coverage by offloading the weak V2I links to high quality V2V links for the cell edge users. The cumulative distribution function (CDF) of the duration of playback video buffered at the receiver is plotted in Fig. \ref{rx_buff}. The Performance of the proposed and baseline methods is compared for different values of reliability ($\epsilon$), resources block ($M$) and vehicle densities ($V$). A neighborhood size of 10 m is assumed in the proposed network slicing. From Fig. \ref{rx_buff} it can be seen that the proposed network slicing algorithm improves the performance of vehicular devices compared to both baselines. When network slicing is used with $\epsilon = 0.1$ we see that the buffered video is greater than the target threshold for all vehicles. Moreover, when we increase the reliability i.e. $\epsilon = 0.01$, the number of vehicles achieving the target threshold increases along with an increase in buffered video per vehicle. On the other hand, when we increase resources while keeping the same value of reliability higher numbers of chunks are available at the receiver buffer to be watched compared to the case with fewer resources. Similar trend can be observed in the baseline methods, where high value of reliability results in more vehicles satisfying the reliability criteria and vice versa. Moreover, the effect of available resources in the baseline scenarios are similar to the observed performance gains of the proposed approach. When we increase the number of resources in the baseline scenarios we see an increase in the number of users satisfying the reliability criteria where, baseline 2 performs the best because of the increased number of vehicular users (especially at the cell-edge) utilizing high quality links.}

\begin{table}
		\centering
		\caption{Simulator parameters.}%
		\label{params}
		\begin{tabular}{p{3.5cm} p{4cm} }
			 \hline
			\textbf{Parameter} & \textbf{Assumption} \\  \hline \hline
			Duplex mode & FDD \\  
			System bandwidth (MHz) & 5  \\  
			& V2I: \hspace{2.5pt}2  \\
			\vspace{-10pt}Carrier frequency (GHz) & V2V: 5.9 \cite{3GPP_frequency} \\  
			Antenna configuration & 1 Tx $\times$ 2 Rx\\ 
			Receiver type & Maximum ratio combining \\  
			Vehicle speed (km/h) & 140  \\  
			& V2I: \hspace{2pt}46  \\
			\vspace{-10pt}	Transmission power (dBm) &	V2V: 20 \\  
			L2S interface metric & MIESM \\  
			Synchronization & Time and frequency synchronized \\  
			HARQ & Chase combining \\  
			Inter RSU distance (m) & 1732 \\  
            Reliability '$\epsilon$' & 0.1, 0.01\\  
            Video qualities & 240p, 360p, 720p \\  
            Rate required (kbps) & 400, 800, 1200 \\  \hline
		\end{tabular}
	\end{table}

\begin{figure}[hbtp]
	\centering
	\includegraphics[width=0.5\textwidth]{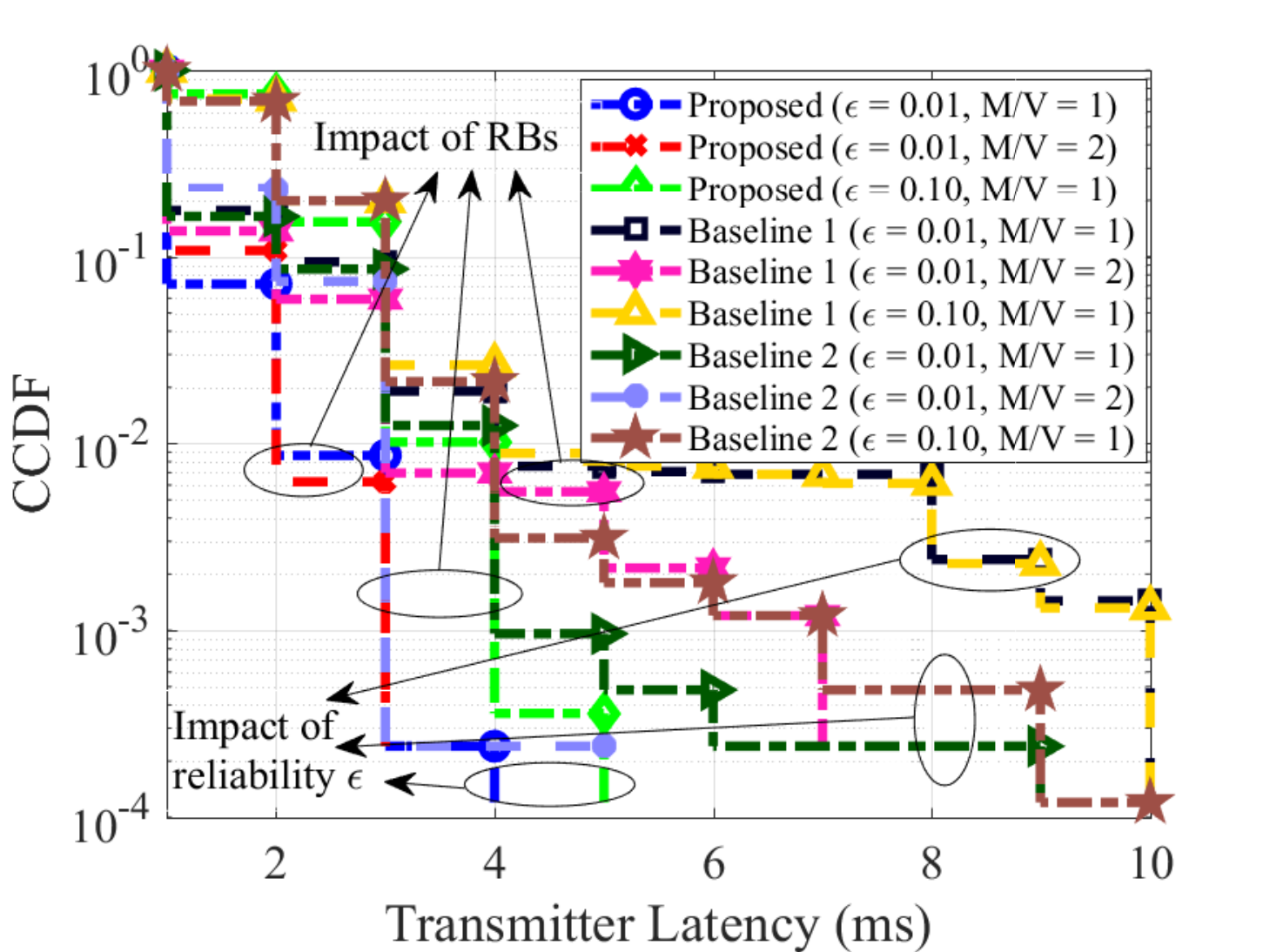}
	\caption{\label{latency}CCDF of queuing latency at the transmitter.}
\end{figure}

\textcolor{black}
{
Since vehicular devices are streaming videos the QoE measures the user satisfaction. In this work the QoE is defined in terms of seamless video experience and quality fluctuations. Seamless video experience is ensured when packets are sent without delay to the receiver. Fig. \ref{latency} shows the complementary cumulative distribution function (CCDF) of the packets queuing latency. From Fig. \ref{latency} we can see that in baseline 1 we have maximum transmitter queuing latency of 10 ms when the resources are equal to the number of users and the latency is reduced to 7 ms when the resources are doubled. In baseline 2 where, the cell-edge vehicles are offloaded to be served by neighboring vehicle performs better than baseline 1. The maximum transmitter latency of 10 ms is observed for lower value of reliability and the latency reduces to 9 ms when the reliability is increased. Similarly, when the number of resources are increased for baseline 2 the latency is reduced to 5 ms. On the other hand, the proposed solution reduces the maximum latency of baselines by half e.g. 5 ms for less reliable scenario i.e. $\epsilon = 0.1$, moreover the latency is further reduced by 1 ms when the reliability is increased. The minimum queuing latency of 3 ms is achieved when network slicing is introduced with double the resources compared to the number of users. Combining the analysis of Fig. \ref{rx_buff} \& \ref{latency} we can see that when the number of resources are greater than the number of users we achieve the minimum queuing latency as well as increased number of users with high buffered amount of playback video still to watch. On the other hand higher reliability increases the QoE of users by minimizing the latency and increasing the duration of playback video available at the receiver and vice versa. It is observed that the baselines method doesn't achieve the reliability criteria for all the vehicles in the network. This is evident from the behavior of receiver buffer in Fig. \ref{rx_buff} and latency in Fig. \ref{latency} where, we can see a fraction of vehicles satisfying the reliability constraints.}

\textcolor{black}
{
In baseline 1 vehicles have no other option than to directly connect to the RSU and receive video streams, but users with bad channels experience erroneous transmissions which leads to increased queuing latency due to retransmissions. Whereas baseline 2 performs better than baseline 1, since cell-edge vehicle with weak links are offloaded to the neighboring vehicles resulting in improved link quality. Baseline 2 doesn't consider the possibility that vehicle other than cell-edge can also possess weak V2I links due to high mobility. The proposed network slicing approach on the other hand decreases the possibility of erroneous transmissions by establishing network wide high quality V2I and V2V links. Vehicles with poor channel have the option to either connect to the RSU or the SL for receiving video streams.
}

\begin{figure}[hbtp]
	\centering
	\includegraphics[width=0.5\textwidth]{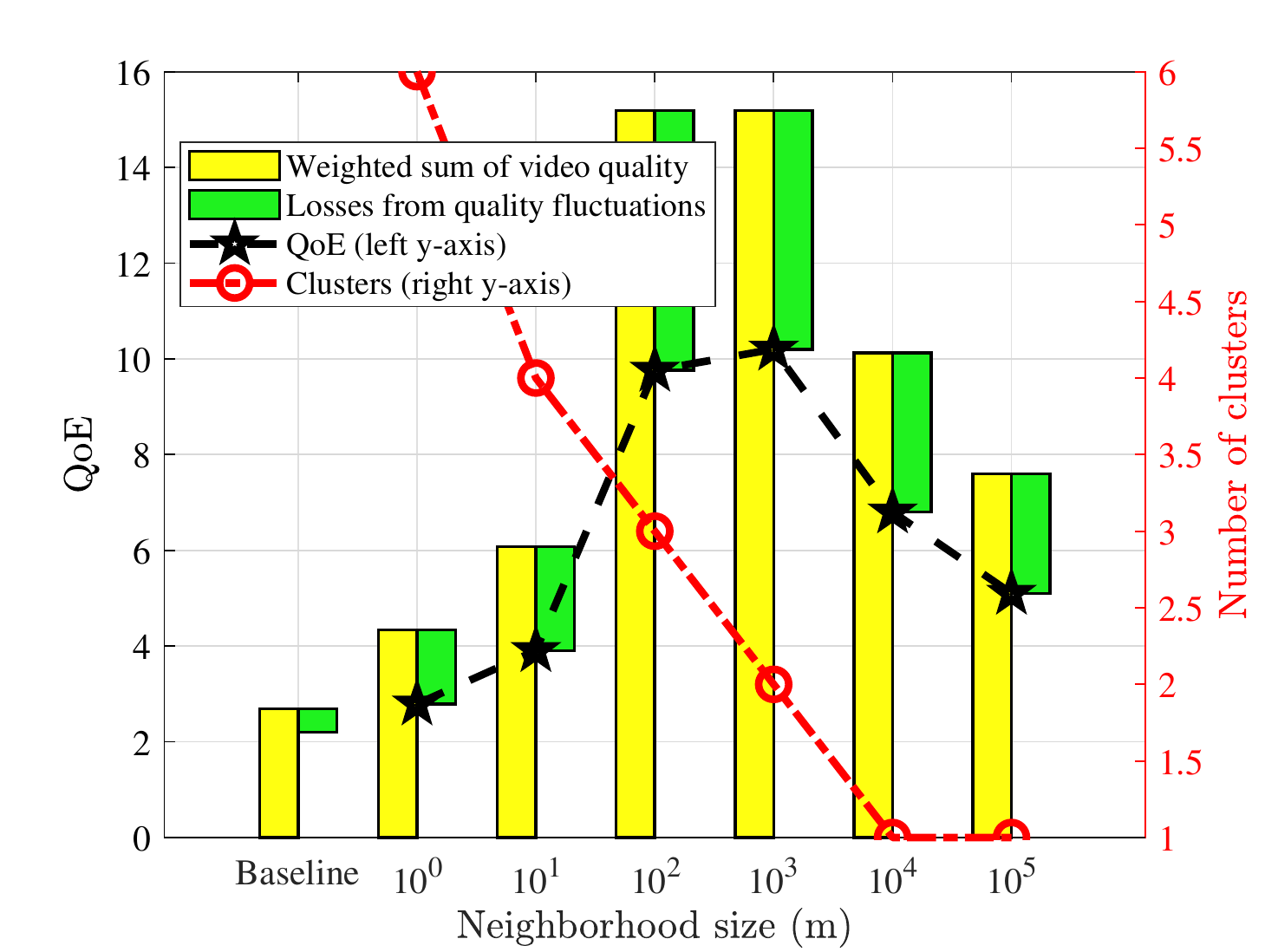}
	\caption{\label{neighborhood}Effect of neighborhood size on QoE.}
\end{figure}

Next, we analyze the effect of neighborhood size on the achieved QoE (left y-axis, black dashed line) and the effect on the number of clusters (right y-axis, red dash-dotted line). Fig. \ref{neighborhood} shows the QoE and the number of clusters for a given neighborhood size. As explained earlier neighborhood size effects the similarity of vehicles i.e. small value of neighborhood size results into more clusters with less number of vehicles in it. We can see from Fig. \ref{neighborhood} that when $\sigma = 10^0 \hspace{2pt} \text{m}$ we have 7 clusters in the network and the number of clusters reduces to 1 when we increase the neighborhood size $\sigma = 10^4 \hspace{2pt}  \text{m}$. When the number of cluster in the network increases the inter cluster interference lowers the performance of V2V links, which results in low value of QoE. On the other extreme when there is only one cluster in the network then slice leader may become overloaded due to large number of free users and in this case vehicles which are not served lowers the QoE. The achieved QoE is maximum when the neighborhood size is between $10^2 \text{-} 10^3 \hspace{2pt} (\text{m})$ when there are 2 and 3 clusters in the network. It should be noted that neighborhood size has direct effect on the performance of network slicing, large neighborhood size in dense networks leads to overloaded clusters, where cluster leaders may not be able to serve their free vehicle and small neighborhood size increases the number of clusters and the inter cluster interference. So neighborhood size should always be chosen based on the density/sparsity of network. In baseline 1 RSU serves all the vehicle and its performance is lower than network slicing, which is because RSU doesn't have high quality V2I links with all the vehicles which leads to lower QoE.

\begin{figure}[hbtp]
	\centering
	\includegraphics[width=0.5\textwidth]{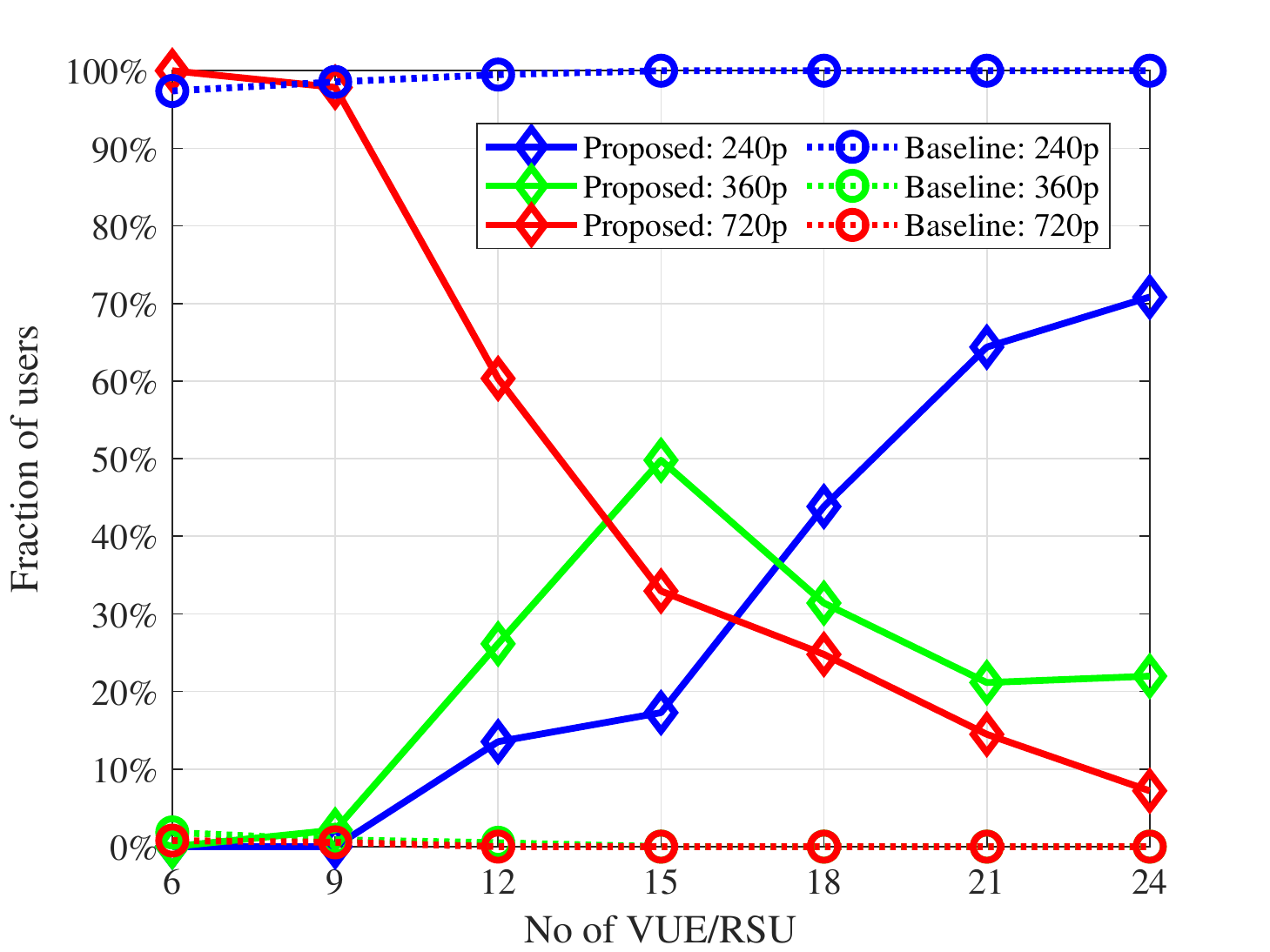}
	\caption{\label{quality_plot}Comparison of the achievable video qualities of the proposed and baseline methods as a function of the number of vehicles per RSU.}
\end{figure}

The video quality selection algorithm is based on the solution of the optimization problem (23), where the optimization variables are the choice of video quality and the resource block assignment. Fig. \ref{quality_plot} shows the achieved video quality for the proposed and baseline 1 as a function of number of vehicular users (VUEs) per RSU. The small number of vehicles per RSU (i.e., 6 VUE/RSU) achieves high quality (720p) video for all the vehicles in the proposed scenario compared to the baseline 1 technique, which achieves low quality (240p) video for $98$\% of the vehicles and the remaining $2$\% VUE achieve higher video quality i.e., 360p, 720p. The proposed technique ensures high quality video experience for increased number of vehicles per RSU compared to baseline 1 because the proposed technique utilizes high quality V2V links along with the V2I links, while baseline 1 only utilizes V2I links. When the number of VUE per RSU is increased to 15, baseline 1 achieves the lowest video quality for $100$\% of the users, while the proposed solution realizes lowest video quality (240p) for $18$\% of the VUEs, medium video quality (360p) for $50$\% of the VUEs, and the highest video quality (720p) for the remaining $32$\% of the users. Further increasing the number of VUEs per RSU results in lower quality video selection and a decline in high quality video experience, which is due to the limited radio resources in the network. The decline in high quality video experience in the proposed case for higher number of VUEs is due to the fact that the Lyapunov video selection algorithm selects the video quality and radio resources to ensure long term stability of all vehicles and to avoid the penalty of video quality fluctuation. Since, the limited number of radio resources cannot provide high quality video experience to all the vehicles, the Lyapunov video selection algorithm lowers the video quality of certain vehicles to avoid the associated cost of quality fluctuation.

%% file: conclusion.tex
In this paper, we have formulated the problem of joint resource allocation and video selection for vehicular devices as a stochastic optimization problem under constraints of queue stability and QoE. The aim is to maximize QoE for vehicular devices while ensuring a  threshold duration of video at each receiver. We simulated a highway scenario with different vehicular densities and analyzed the performance of the proposed technique. Under mild assumptions, we analyzed the performance of the proposed technique with the system level simulator and we have shown that performance of proposed method is better than the baselines. Furthermore, the proposed approach is evaluated for different neighborhood sizes and the effect of changing reliability or resources is compared with the baselines. The considerable gains in the proposed technique is due to the utilization of high quality links which maximized the user's experience. 

%% file: ack.tex
This research was supported by the Kvantum institute strategic project SAFARI, High5 project number 2192/31/2016 funded by Business Finland, Bittium, Keysight, Kyynel, MediaTek, Nokia, University of Oulu and the Academy of Finland 6Genesis Flagship project under grant 318927. First author is grateful for the grants received from the HPY, Nokia, and Walter Ahlströmin Foundation. We would like to thank Dr. Anis Elgabli for helping in the revision process of this work.  